\documentclass[sigconf]{acmart}

\usepackage{booktabs} 

\usepackage{hyperref}
\setcopyright{acmlicensed}

\copyrightyear{2024}
\acmYear{2024}
\setcopyright{rightsretained}
\acmConference[ICVGIP 2024]{Indian Conference on Computer Vision Graphics and Image Processing}{December 13--15, 2024}{Bengaluru, India}
\acmBooktitle{Indian Conference on Computer Vision Graphics and Image Processing (ICVGIP 2024), December 13--15, 2024, Bengaluru, India}
\acmPrice{}
\acmDOI{10.1145/3702250.3702284}
\acmISBN{979-8-4007-1075-9/24/12}

\begin{document}
\title{Leveraging Auxiliary Classification for Rib Fracture Segmentation}

\author{Harini G.}
\authornote{Both authors contributed equally to this research.}
\email{m22ma005@iitj.ac.in }
\affiliation{%
  \institution{Indian Institute of Technology Jodhpur}
  \city{Jodhpur}
  \state{Rajasthan}
  \country{India}
}
\author{Aiman Farooq}
\authornotemark[1]
\email{farooq.1@iitj.ac.in}
\affiliation{%
  \institution{Indian Institute of Technology Jodhpur}
  \city{Jodhpur}
  \state{Rajasthan}
  \country{India}
}

\author{Deepak Mishra}

\email{dmishra@iitj.ac.in}
\affiliation{%
  \institution{Indian Institute of Technology Jodhpur}
  \city{Jodhpur}
  \state{Rajasthan}
  \country{India}
}
\renewcommand{\shortauthors}{}

\begin{abstract}
Thoracic trauma often results in rib fractures, which demand swift and accurate diagnosis for effective treatment. However, detecting these fractures on rib CT scans poses considerable challenges, involving the analysis of many image slices in sequence. Despite notable advancements in algorithms for automated fracture segmentation, the persisting challenges stem from the diverse shapes and sizes of these fractures. To address these issues, this study introduces a sophisticated deep-learning model with an auxiliary classification task designed to enhance the accuracy of rib fracture segmentation. The auxiliary classification task is crucial in distinguishing between fractured ribs and negative regions, encompassing non-fractured ribs and surrounding tissues, from the patches obtained from CT scans. By leveraging this auxiliary task, the model aims to improve feature representation at the bottleneck layer by highlighting the regions of interest. Experimental results on the RibFrac dataset demonstrate significant improvement in segmentation performance. 
\end{abstract}

%
%





\maketitle

\section{Introduction}

\begin{figure}
    \centering
    \includegraphics [width=\linewidth]{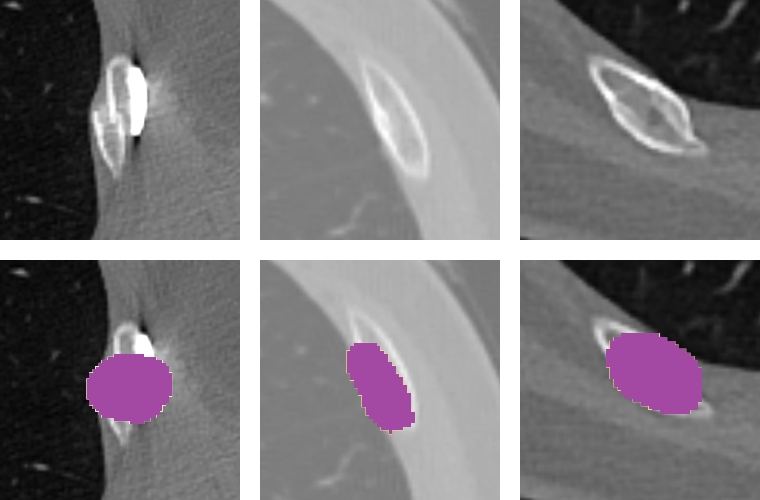}
    \caption{The first row of images displays fractured regions of different sizes and shapes, appearing as irregularities and disruptions in the bone structure. In the second row, the ground truth annotations are overlaid on the fractured regions in the corresponding CT images from the first row. }
    \label{fig:model}
\end{figure}

Rib fractures present a critical concern in cases of traumatic chest injuries due to their pivotal role in protecting the thoracic organs \cite{dogrul2020blunt}. These fractures, often stemming from blunt force trauma, can vary widely in severity from minor cracks to complete breaks, each carrying significant implications for patient health. When ribs sustain fractures, especially in multiple locations, they can give rise to a range of complications, raising immediate concerns such as the potential for thoracic hemorrhage \cite{kuo2019rib} and pneumothorax \cite{kim2020chest}. Additionally, the fractured ends of the ribs harbor the potential to inflict harm upon adjacent structures, including the heart, lungs, or major blood vessels \cite{davoodabadi2022correlation}, underscoring the critical need for prompt and thorough medical assessment and treatment. Through the precise diagnosis and management of rib fractures, healthcare teams can mitigate potential complications, minimize associated risks, and enhance overall patient care in the realm of traumatic chest injuries. Computed Tomography (CT) scans have emerged as the preferred modality for diagnosing rib fractures and assessing accompanying injuries in patients presenting with traumatic chest injuries \cite{sochor20033d,oikonomou2011ct,awais2019diagnostic}. CT scans offer highly detailed, cross-sectional images of the chest area, enabling healthcare providers to visualize fractures precisely and determine their extent and severity. Despite the remarkable capabilities of CT scans, manually reviewing these CT slices for rib fractures can be labor-intensive and susceptible to human error. This is particularly evident in cases where fractures are subtle or distributed across multiple slices of the scan, as shown in Fig. \ref{fig:model}. A meticulous examination is essential, as undetected or misinterpreted fractures can seriously affect patient care and treatment outcomes. Thus, there is a strong imperative to develop automated and accurate methods for identifying rib fractures in clinical applications.

Traditional methods to segment bone fractures \cite {anu2015detection,ruikar2019segmentation} rely on handcrafted feature extraction followed by segmentation and are highly susceptible to errors. The application of deep learning (DL) in medical imaging has revolutionized medical image segmentation, offering algorithms that streamline the process by automatically extracting intricate features from imaging data. Models such as UNet \cite{ronneberger2015u}, UNet++ \cite{zhou2018unet++}, SegNet \cite{xing2022cm}, and FCN \cite{roth2018application} have become indispensable for semantic segmentation, offering a robust framework for medical image analysis. The encoder-decoder structure of UNet and skip connections make it particularly adept at handling the complexities of medical imaging. It is crucial to consider the unique characteristics of the rib region when segmenting fractures to avoid false positives from distant areas and gain insight into fracture patterns in healthy bone structures. Jin et al. \cite{jin2020deep} proposed FracNet, a 3DUNet-based model for segmenting fractured ribs on the RibFrac dataset. To handle variations in fracture size, Liu et al. \cite{liu2021multi} devised a multi-scale network to integrate multiple sizes of fractures to minimize size alteration. Cao et al. \cite{cao2023robust} proposed SA-FracNet, a fracture shape-aware multi-task segmentation network to delineate the fracture and improve performance over the FracNet model. However, SA-FracNet necessitates an extra network trained through self-supervised contrastive learning for its initialization, which can distinguish between fractured and non-fractured regions.

The investigation by Phan et al.\cite{phan2023exploiting} utilizes an additional segmentation task to improve the performance of image-level classification. Considering an alternative approach, the patch-level annotations can be used to identify the voxel-level annotations. This patch-level classification task facilitates a more thorough understanding of regions where fractures are more probable to occur, thereby serving as an auxiliary task. We propose a network that segments fracture regions aided by the auxiliary classification task, which helps identify the potential fracture sites through a classifier and increases the segmentation performance.
\section{Related Work}

\subsection{Rib Fracture Segmentation}

Numerous methods have been introduced to address the challenging task of segmenting fractures within the intricate rib region from CT scans. Jin et al. \cite{jin2020deep} presented Fracnet, a specialized network utilizing a UNet backbone tailored explicitly for fracture segmentation. This network is further refined with post-processing steps, aiming to enhance the precision of fracture segmentation. In their research, Wu et al. \cite{wu2021development} introduced a comprehensive framework that unifies rib segmentation, fracture segmentation, and fracture localization into a single cohesive system. This joint framework is designed to capitalize on the inherent synergy among these tasks for identifying and delineating fractures within the segmented rib structures. \cite{yang2024deep} has introduced a network containing Fracnet as an internal baseline and also incorporating additional information extracted from the voxel features of the rib structures in the baseline. The intricate and complex nature of rib structures poses a significant challenge for automated fracture detection algorithms. To address this issue, researchers have explored various approaches incorporating rib segmentation, either as a dedicated step or integrated within the fracture detection process. Jin et al. \cite{jin2020deep} introduced rib segmentation as a post-processing measure to reduce false positives occurring distant from the lung parenchyma. This method refines the segmentation process by focusing on the specific characteristics of rib structures, improving the fracture segmentation performance.
In contrast, Wu et al. \cite{wu2021development} proposed a method where rib segmentation is integrated as part of a joint task with fracture detection and localization, ensuring that the rib structures are accurately delineated for better contextual understanding. Incorporating rib segmentation into a unified framework enhances fracture identification and precise location within their proprietary dataset. Taking a different approach, Yang et al. \cite{yang2024deep} employed rib segmentation to extract features of ribs and integrated with segmentation in their methodology. Whether implemented before or after the primary fracture segmentation step, rib segmentation is crucial in identifying fractures within CT scans. It introduces additional complexity to the segmentation process, considering the challenges posed by the intricate nature of rib structures. 

\begin{figure*}
    \centering
    \includegraphics [width=14cm, height=5cm]{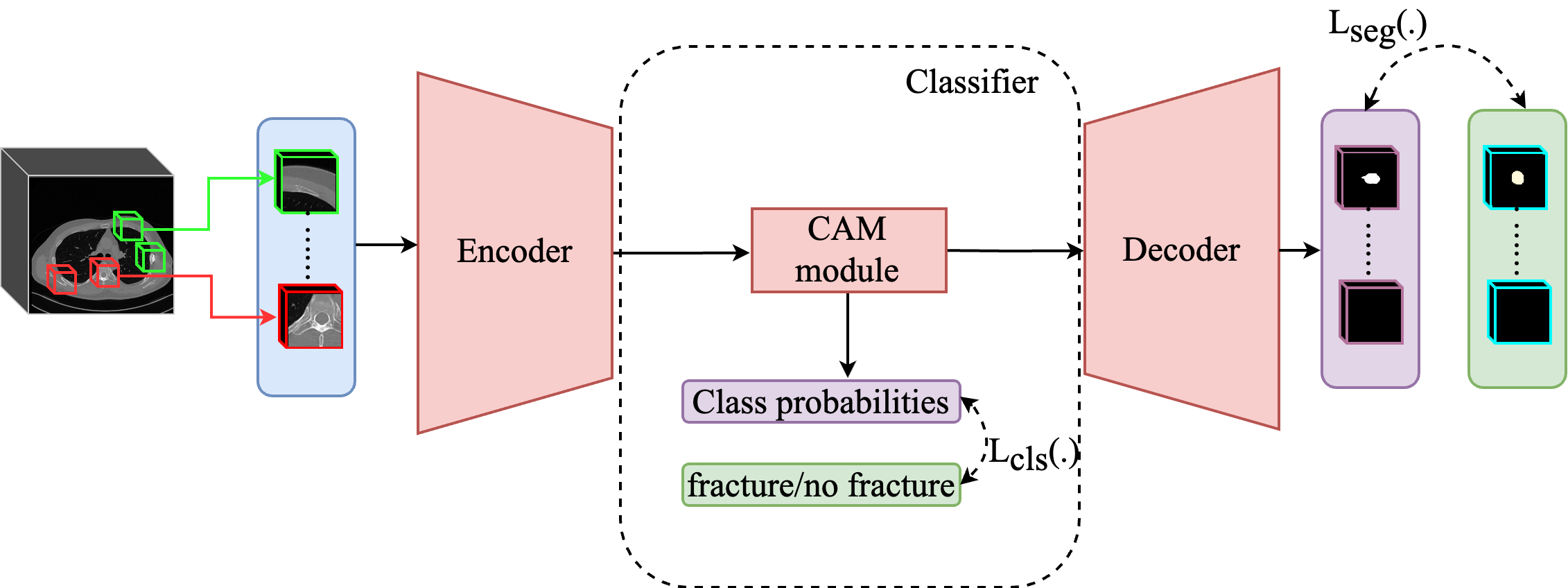}
    \caption{Schematic view of our proposed segmentation network with an auxiliary classifier at the bottleneck.}
    \label{fig:network}
\end{figure*}

\subsection{Weak Supervision}
Various medical imaging applications focused on segmentation often leverage auxiliary tasks to improve outcomes. Studies such as Myronenko et al. \cite{myronenko20183d} and Phan et al. \cite{phan2023exploiting} have demonstrated the significant benefits of coupling an auxiliary task with the primary objective, leading to enhanced effectiveness of networks in medical imaging tasks, regardless of the imaging modality. In the work by Myronenko et al. \cite{myronenko20183d}, a reconstruction task using a variational autoencoder \cite{kingma2013auto} from bottleneck features serves as an auxiliary task that aids in the segmentation of tumors from MRI images. This approach highlights the potential of auxiliary tasks to provide additional information and context to the primary segmentation task, ultimately improving the accuracy and reliability of tumor segmentation. On a similar note, Phan et al. \cite{phan2023exploiting} introduced a weakly supervised segmentation approach that assists in the classification of microplastics from micro-spectroscopy images. By incorporating this auxiliary task, the network can pinpoint important features and regions within the images, enhancing the classification accuracy of microplastics. The segmentation approach plays a crucial role in aiding classification tasks by identifying the active areas within an image, thereby assisting in determining the image class. Class Activation Maps (CAMs) \cite{zhou2016learning} are mainly instrumental in generating activation maps highlighting potential discriminative features within the images. Huang et al.  \cite{huang2018weakly} further demonstrate the generation of segmentation maps, mainly when pixel-level annotations are unavailable but image-level annotations are provided. Their work showcases how these active regions generate segmentation maps, aiding in object localization and classification tasks. 

\subsection{Attention}
Attention is a crucial aspect of human perception for analyzing objects. Various attention modules have been proposed to enhance the importance of discriminative features within images, improving performance on specific tasks. The Convolutional Block Attention Module (CBAM), as introduced in the work by Woo et al. \cite{woo2018cbam}, incorporates both channel and spatial attention modules to capture relationships between channels and spatial positions, respectively. Similarly, the self-attention mechanism of the Vision Transformer \cite{dosovitskiy2020image} is instrumental in learning global relationships across images. This has led to several architectures tailored for segmentation tasks, such as UNETR \cite{hatamizadeh2022unetr} and TransUNet \cite{chen2021transunet}. All these studies have demonstrated incorporating attention mechanisms enhances image perception by effectively modeling intricate relationships within the image for improved segmentation performance.

Based on the findings of these studies, we suggest a model that includes an additional classification task. This task is designed to identify the salient features of patches, which are active while predicting the class label and enhance it in the bottleneck, leading to improved segmentation performance.
\section{Methodology}


The architecture shown in Fig. \ref{fig:network} illustrates our proposed model, designed with an encoder-decoder structure. Featuring a CAM module in the bottleneck, our model classifies the features and produces an activation map, pinpointing regions crucial for determining classification labels. Leveraging these active regions, the segmentation network receives added guidance for improved performance.

\subsection{Pre processing}
In Fig. \ref{fig:network}, we present our method of extracting patches of size 128x128x128 from CT scans to represent positive and negative instances. Positive patches are centered around the centroid of the fracture region and undergo random translations for augmentation. If multiple fractures are within the patch size, separate patches are extracted around the centroid of each fracture, potentially causing overlap. Random translations are used to avoid encountering the fractures always in the center of the patch. Conversely, negative patches are sampled from the entire CT volume, excluding the fracture region. Specifically, we obtain one-third of the negative regions from the spine region, another one-third by identifying and cropping around the symmetric counterpart of the positive centroid, and the final third is randomly selected from the complete CT volume. The negative patches are chosen so they do not contain any voxels of the positive region. Importantly, we ensure equal patches for both fracture and non-fracture regions. These selected patches then undergo a series of morphological operations, such as thresholding, to isolate voxels within the intensity range of -200 to 1000. A normalization step is implemented to standardize the voxel intensities, scaling them from -1 to 1. To reduce the training time, all the patches are extracted, preprocessed, and saved in memory to reduce the processing time. While inferring the patches are extracted from all possible centroids in CT volume.

\subsection{Network architecture}

As depicted in Fig. \ref{fig:network}, the neural network processes a batch of 3D patches within its encoder branch, adhering to the standard UNet architecture with channel sizes \{16, 32, 64, 128\} and each computes feature representations $ \text{e}_\text{{li}} $ representing layer l of $\text{i}^\text{{th}}$ element in a batch and $\text{l} \in \{\text{0}, \text{1}, \text{2}, \text{3}\}$. $ \text{e}_\text{{0i}} $ represents input to the first encoder block that is a pre-processed patch, and the final encoder output is $ \text{e}_\text{{3i}}$. The bottleneck features $ \text{e}_\text{{3i}}$ serve as input to the CAM module, which is crucial in guiding the segmentation task by determining whether a patch contains a fracture.

\begin{figure}
    \centering
    \includegraphics [width=0.7\linewidth]{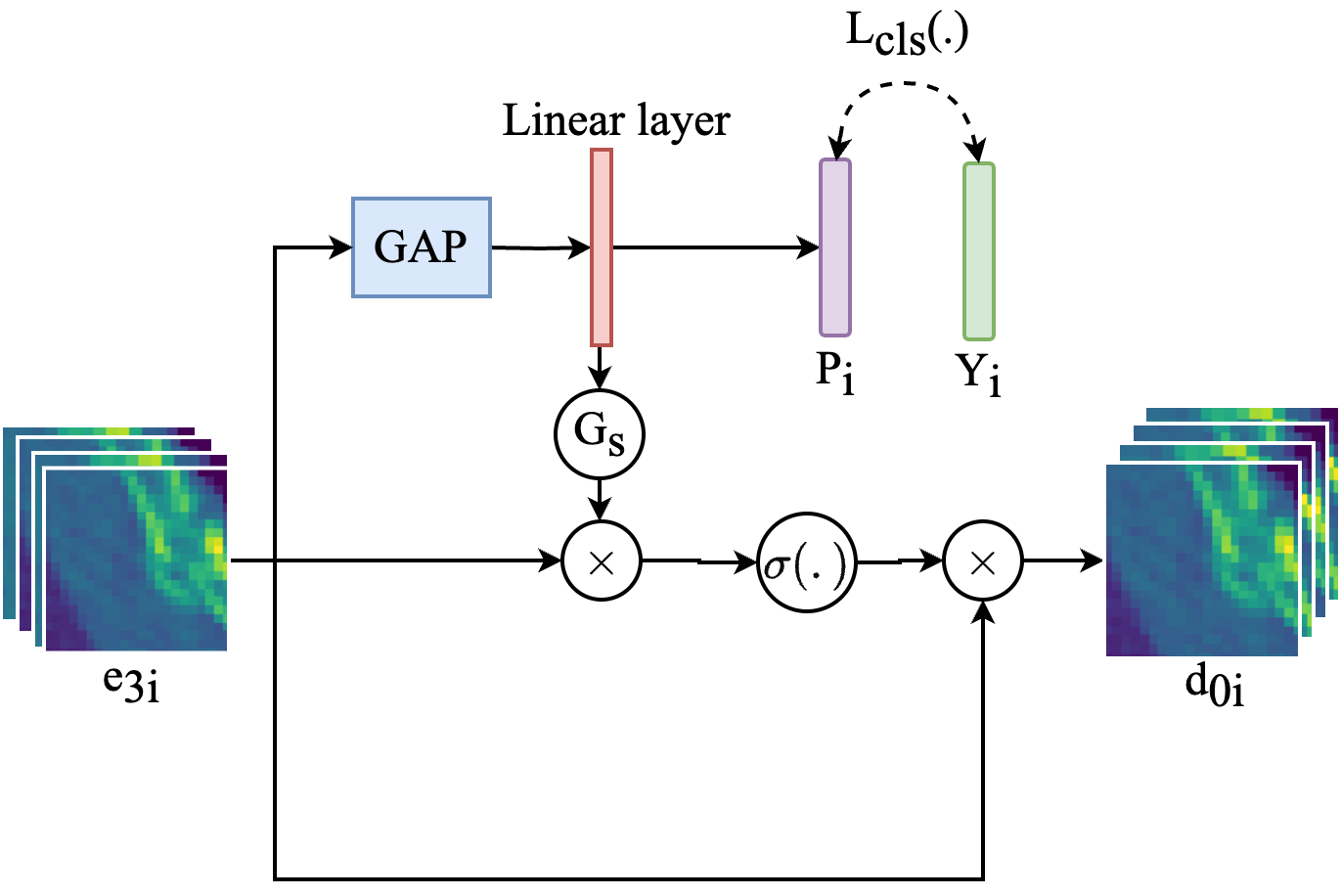}
    \caption{Schematic view of CAM module.}
    \label{fig:cam}
\end{figure}

\textbf{CAM module} : As shown in Fig. \ref{fig:cam}, the features $ \text{e}_\text{{3i}}$  are subjected to global average pooling (GAP) along the channel dimension and serve as input to a linear layer to predict whether a patch contains a fracture or not. Subsequently, the obtained weights from the linear layer are multiplied with each channel of $ \text{e}_\text{{3i}}$  and passed through a sigmoid function. The resulting value is then multiplied with the final layer encoder features, mathematically shown by the following equations.
\begin{equation}
\text{F}_{\text{ci}} = \text{GAP}(\text{e}_{\text{3i}}) = \frac{1}{\text{W} \times \text{H} \times \text{D}}  \sum_{\text{l}=1}^{\text{W}}  \sum_{\text{m}=1}^{\text{H}} \sum_{\text{n}=1}^{\text{D}} (\text{e}_{\text{3i}})\label{eq:Fci}
\end{equation}

Where $\text{F}_{\text{ci}}$ is a resultant tensor, $\text{GAP(.)}$ represent the Global Average Pooling operation along the channel axis and $\text{W}$, $\text{H}$, and $\text{D}$ denote the width, height, and depth of $ \text{e}_\text{{3i}} $, respectively and $\text{i}$ represents the  $\text{i}^{\text{th}}$ sample of a batch and c is number of channels in $\text{e}_\text{{3i}}$.
\begin{equation}
\text{P}_{\text{i}} = \sigma({\text{W}_{\text{ci}}}^{T}.\text{F}_{\text{ci}})
\end{equation}
Where, $\text{W}_{\text{ci}}$ denotes the weight matrix where $\text{c}$ is the number of channels in $\text{e}_\text{{3i}}$, $\text{F}_\text{{ci}}$ denotes a tensor computed using equation \ref{eq:Fci} of $\text{i}^\text{th}$ sample in batch B, $T$ represents transpose operation, $\sigma(.)$ denotes sigmoid operation, and $\text{P}_\text{i}$ represents the probability of a patch containing a fracture.
\begin{equation}
    \text{d}_{\text{0i}} = \text{e}_{\text{3i}} \circ \sigma(G_{S}(\text{W}_{\text{ci}})\times(\text{e}_{\text{3i}}))
\end{equation}

Where operations $G_S$ and $\circ$ represent re-expansion and element-wise multiplication and $\text{d}_{\text{0i}}$ represents the final computed features that act as input to further layers.

The computed features $\text{d}_{\text{0i}}$ are input to the decoder branch. At each layer, features $\text{d}_{\text{li}}$  are computed with the same channel numbers as those in the encoder branch, originating from the bottleneck where $\text{l} \in \{\text{0}, \text{1}, \text{2}, \text{3}\}$. The final features $d_{3i}$ are passed through a convolutional layer and then the sigmoid function to obtain the mask.

\textbf{Loss function}: The weights are updated based on the difference between the predicted mask and the target mask, as well as the class probabilities predicted at the bottleneck concerning the target class.

\begin{equation}
 \mathcal{L}_{\text{total}} =  \mathcal{L}_{\text{seg}}  + \theta(\tau).\mathcal{L}_{\text{cls}}
\end{equation}
Where \( \mathcal{L}_{\text{seg}} \), \( \mathcal{L}_{\text{cls}} \) represent the segmentation loss, classification loss, and $\theta(\tau)$ represents a monotonically decreasing function with respect to $\tau$, where $\tau$ represents epoch, thereby ensuring that minor errors in classification do not adversely affect the segmentation performance in later epochs.
\begin{equation}
 \mathcal{L}_{\text{seg}} =  \alpha_{\text{1}}.\mathcal{L}_{\text{focal}}  + \alpha_{\text{2}}.\mathcal{L}_{\text{dice}}
\end{equation}
Where \( \mathcal{L}_{\text{focal}} \), \( \mathcal{L}_{\text{dice}} \) represent focal loss and dice loss respectively and $\alpha_{\text{1}}$, $\alpha_{\text{2}}$ are hyperparameters.
\begin{equation}
\mathcal{L}_{\text{focal}}(\text{p}_{\text{it}})= -\frac{1}{\text{BN}}\sum_{\text{i=1}}^{\text{B}}\sum_{\text{t=1}}^{\text{N}}((\text{1-p}_{\text{it}})^\gamma log(\text{p}_{\text{it}}))
\end{equation}
\[
\begin{aligned}
    \text{p}_{\text{it}} &= 
    \begin{cases}
        \text{1-p}_{\text{iv}}, & \text{if } \text{y}_{\text{iv}} = 1 \\
        \text{p}_{\text{iv}}, & \text{otherwise}
    \end{cases}
\end{aligned}
\]

Where $\text{p}_{\text{iv}}$  is the probability of the $\text{v}^\text{{th}}$  voxel being a fractured region from $\text{i}^\text{{th}}$ patch in a batch of size B, $\text{y}_{\text{iv}}$ is the target label of $\text{v}^\text{{th}}$  voxel, and $v \in \{1, N\}$, where $\text{N} = \text{W} \times \text{H} \times \text{D} $ and $\gamma$ is a constant.

\begin{equation}
\mathcal{L}_{\text{dice}} = \frac{1}{\text{B}}\sum_{\text{i=1}}^{\text{B}} (1 - \frac{2 \sum_{\text{v=1}}^{\text{N}} \text{g}_\text{iv}\text{p}_\text{iv} + \epsilon}{ \sum_{\text{v=1}}^{\text{N}} \text{\text{g}}_{\text{iv}} + \sum_{\text{v=1}}^{\text{N}} \text{p}_\text{iv} + \epsilon})
\end{equation}

Where $\text{g}_\text{iv}$, $\text{p}_\text{iv}$ represents ground truth and prediction probability of $\text{v}^\text{{th}}$  voxel of $\text{i}^\text{{th}}$ patch in a batch $\text{B}$ and $\epsilon$ represents a constant.

Classification loss used is binary cross entropy loss predicting whether a patch contains fracture or not is given by the following equation.
\begin{equation}
\mathcal{L}_{\text{cls}} = -\frac{1}{\text{B}}\sum_{\text{i=1}}^{\text{B}} (\text{Y}_{\text{i}}log(\text{P}_{\text{i}}) + \text{(1 - Y}_{\text{i}})log(\text{1 - P}_{\text{i}}))
\end{equation}

Where $\text{Y}_{\text{i}} $ and $\text{P}_{\text{i}}$ are the target label and predicted class probability of $\text{i}^\text{{th}}$ patch of batch size B.

\subsection{Post processing}
The final prediction is determined by taking the mean of the predictions. Each voxel is then classified as a positive or negative outcome based on a specified probability threshold. Subsequently, the optimistic predictions are grouped into connected components. These connected components are regarded as segmentation proposals, with their probability calculated by averaging the original segmentation scores across all components' voxels. Predictions below a particular size area are excluded to reduce false positives. Moreover, predictions from voxels in the spine region and those with an intensity lower than that of bone are also discarded.
\section{Experiments}
This section summarizes the details of the dataset, experimental settings, and results.

\subsection{Dataset}

The RibFrac Dataset \cite{ribfracchallenge2024,ribfracclinical2020}, featured in the challenge, has been meticulously curated with a specific emphasis on segmenting and classifying rib fractures. This publicly available dataset offers a comprehensive collection of chest-abdomen CT scans and detailed annotations dedicated to rib fracture segmentation and classification across a cohort of 660 patients.

\begin{table*}[htbp]
    \centering
    \caption{Comparison of FROC and Dice scores for different segmentation backbones for rib fracture segmentation. The best results are denoted in blue, whereas the subsequent closest results are highlighted in red.}
    \begin{tabular}{l *{6}{c}}
        \toprule
        Backbone       & \multicolumn{5}{c}{FROC}                   & DSC (in \%)\\
                       & 1      & 2      & 4      & 8      & 0.5    &      \\
        \midrule
        UNETR \cite{hatamizadeh2022unetr}          &  $0.49_{\pm 0.02}$ & $0.56_{\pm 0.02}$ & $0.60_{\pm 0.02}$ & $0.49_{\pm 0.02}$ & $0.63_{\pm 0.02}$ & $48.57_{\pm 2.65}$ \\
        UNet \cite{ronneberger2015u}               & $0.60_{_\pm 0.08}$ & $0.65_{\pm 0.07}$ & $0.67_{\pm 0.08}$ & $0.67_{\pm 0.08}$ & $0.53_{\pm 0.08}$ & $53.00_{\pm 6.07}$ \\
        TransUNet \cite{chen2021transunet}         & $0.63_{\pm 0.02}$ & $0.74_{\pm 0.01}$ & $0.79_{\pm 0.01}$ & $0.80_{\pm 0.01}$ & $0.50_{\pm 0.02}$ & $61.82_{\pm 1.46}$ \\
        AttentionUNet \cite{oktay2018attention}   & $0.69_{\pm 0.00}$ & $0.76_{\pm 0.01}$ & $0.78_{\pm 0.01}$ & $0.78_{\pm 0.01}$ & $0.58_{\pm 0.01}$ & $61.85_{\pm 1.86}$ \\
        FracNet \cite{jin2020deep}  & {$0.69_{\pm 0.02}$} & \textcolor{red}{$0.77_{\pm 0.01}$} & \textcolor{red}{$0.80_{\pm 0.01}$} & \textcolor{red}{$0.80_{\pm 0.01}$} & {$0.56_{\pm 0.04}$} &  \textcolor{red}{$63.38_{\pm 1.63}$} \\
        \midrule
         $\textbf{Ours}_{\textbf{(without classifier)}}$ & \textcolor{red}{$0.71_{\pm 0.03}$} & {$0.75_{\pm 0.01}$} & {$0.78_{\pm 0.01}$} & {$0.78_{\pm 0.01}$} & \textcolor{blue}{$0.60_{\pm 0.01}$} & {$62.21_{\pm 0.45}$} \\

        $\textbf{Ours}_{\textbf{(with classifier)}}$                             & \textcolor{blue}{$0.71_{\pm 0.03}$} & \textcolor{blue}{$0.79_{\pm 0.01}$} & \textcolor{blue}{$0.81_{\pm 0.00}$} & \textcolor{blue}{$0.81_{\pm 0.00}$} & \textcolor{red}{$0.58_{\pm 0.03}$} & \textcolor{blue}{$64.55_{\pm 0.45}$} \\
        \bottomrule
    \end{tabular}
    \label{tab:results1}
\end{table*}
\begin{figure*}[]
    \centering
    \includegraphics [width=\linewidth, height=12cm]{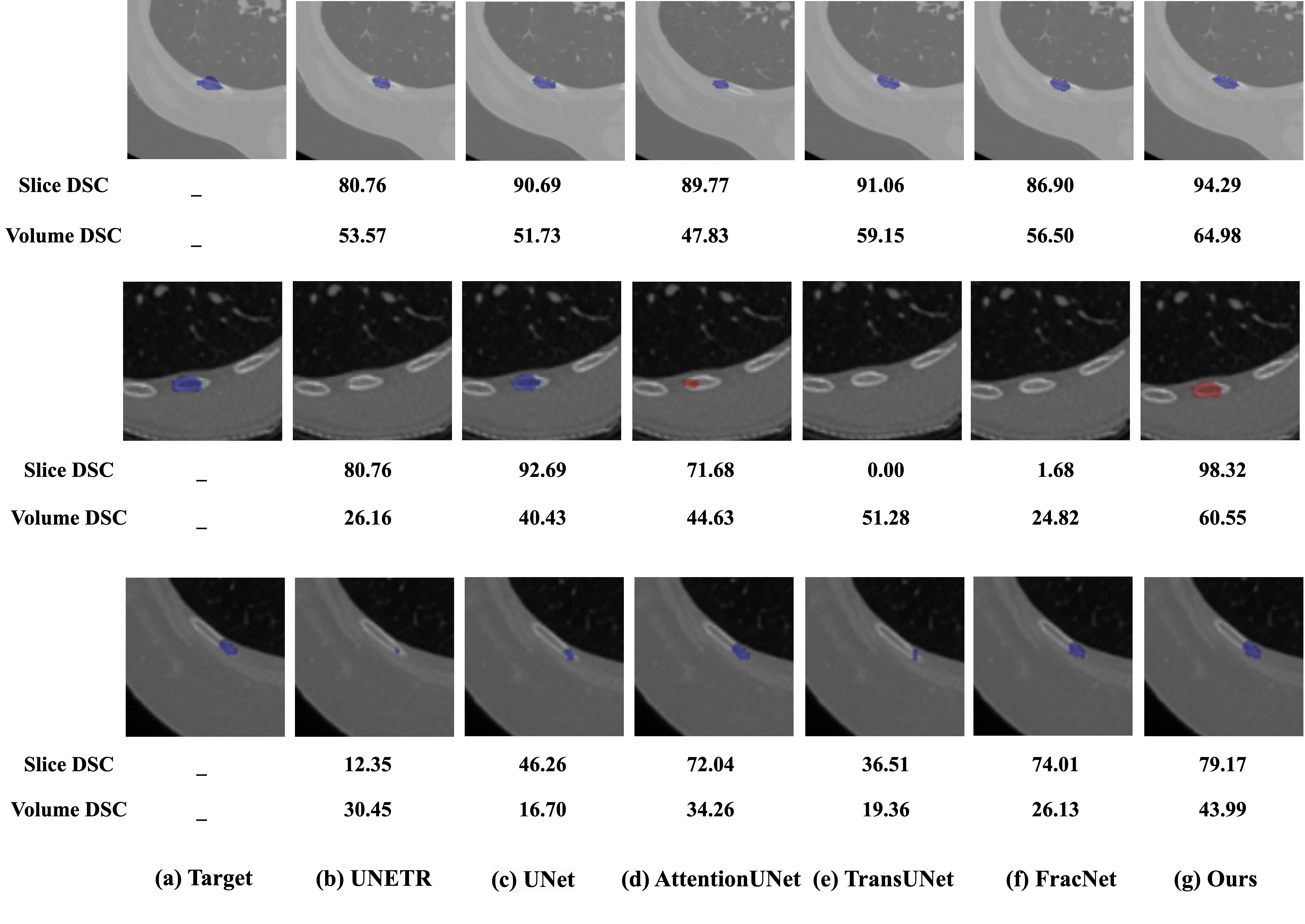}
    \caption{Baseline Models Comparison with volume and slice DSC: First row: Segmentation of Sample 431, Slice Number 268, Second row: Segmentation of Sample 440, Slice Number 159, Third row: Segmentation of Sample 464, Slice Number 278.}
    \label{fig:results}
\end{figure*}

The primary objective of this challenge is to propel advancements in medical image analysis by fostering the creation and validation of novel algorithms designed for rib fracture segmentation and classification. The dataset encompasses approximately 5,000 rib fractures, comprising 420 CT scans for training purposes, each containing a minimum of two fractures. Additionally, there are 80 CT scans reserved for validation and an extra 160 for thorough evaluation testing. Notably, the volumes within the validation and test sets may or may not include fractures.

\subsection{Experimental Settings and Evaluation metrics}

The model was initialized using Kaiming Normal (He initialization) \cite{he2015delving}. All trainable parameters in the image encoder, decoder, and classification networks were updated during training. The loss function is a weighted combination of dice loss and focal loss for segmentation and binary cross-entropy loss for classification. The training utilized the Adam optimizer with parameters $\beta_1 = 0.9$ and $\beta_2 = 0.999$, an initial learning rate 1e-3, and weight decay of 0.01. A global batch size 16 was used for training on an NVIDIA GeForce RTX 3090 (2 GPUs, 24GB each). The training spanned 100 epochs, with model selection based on achieving the minimum validation loss. Each baseline was trained three times with random initialization.

\begin{table*}[htbp]
    \centering
    \caption{Comparison of FROC and Dice scores for different probability and size thresholds. The best results are denoted in blue, whereas the subsequent closest results are highlighted in red.}
    \begin{tabular}{c @{\hspace{10pt}} c *{5}{c} c}
        \toprule
        Probability Threshold & Size Threshold & \multicolumn{5}{c}{FROC} & DSC (in \%) \\
                              &                & 1      & 2      & 4      & 8      & 0.5    &      \\
        \midrule
        0.4      & 150      & 0.72 & 0.78 & 0.81 & 0.82 & 0.81 & 63.90 \\
        0.5      & 150      & 0.72 & 0.78 & 0.81 & 0.81 & 0.59 & 64.18 \\
        0.6      & 50       & 0.71 & 0.78 & 0.80 & 0.81 & 0.57 & 64.49 \\
        0.6      & 100      & 0.71 & 0.78 & 0.81 & 0.81 & 0.57 & 64.52 \\
        \textcolor{red}{0.6}      & \textcolor{red}{200}      & \textcolor{red}{0.71} & \textcolor{red}{0.79} & \textcolor{red}{0.80} & \textcolor{red}{0.81} & \textcolor{red}{0.57} & \textcolor{red}{64.53} \\
        \textcolor{blue}{0.6}       & \textcolor{blue}{150}       & \textcolor{blue}{0.71} & \textcolor{blue}{0.79} & \textcolor{blue}{0.81} & \textcolor{blue}{0.81} & \textcolor{blue}{0.58} & \textcolor{blue}{64.55} \\
        \bottomrule
    \end{tabular}
    \label{tab:results2}
\end{table*}
\subsection{Results}
\begin{figure}[]
    \centering
    \includegraphics [width=0.8\linewidth, height=6cm]{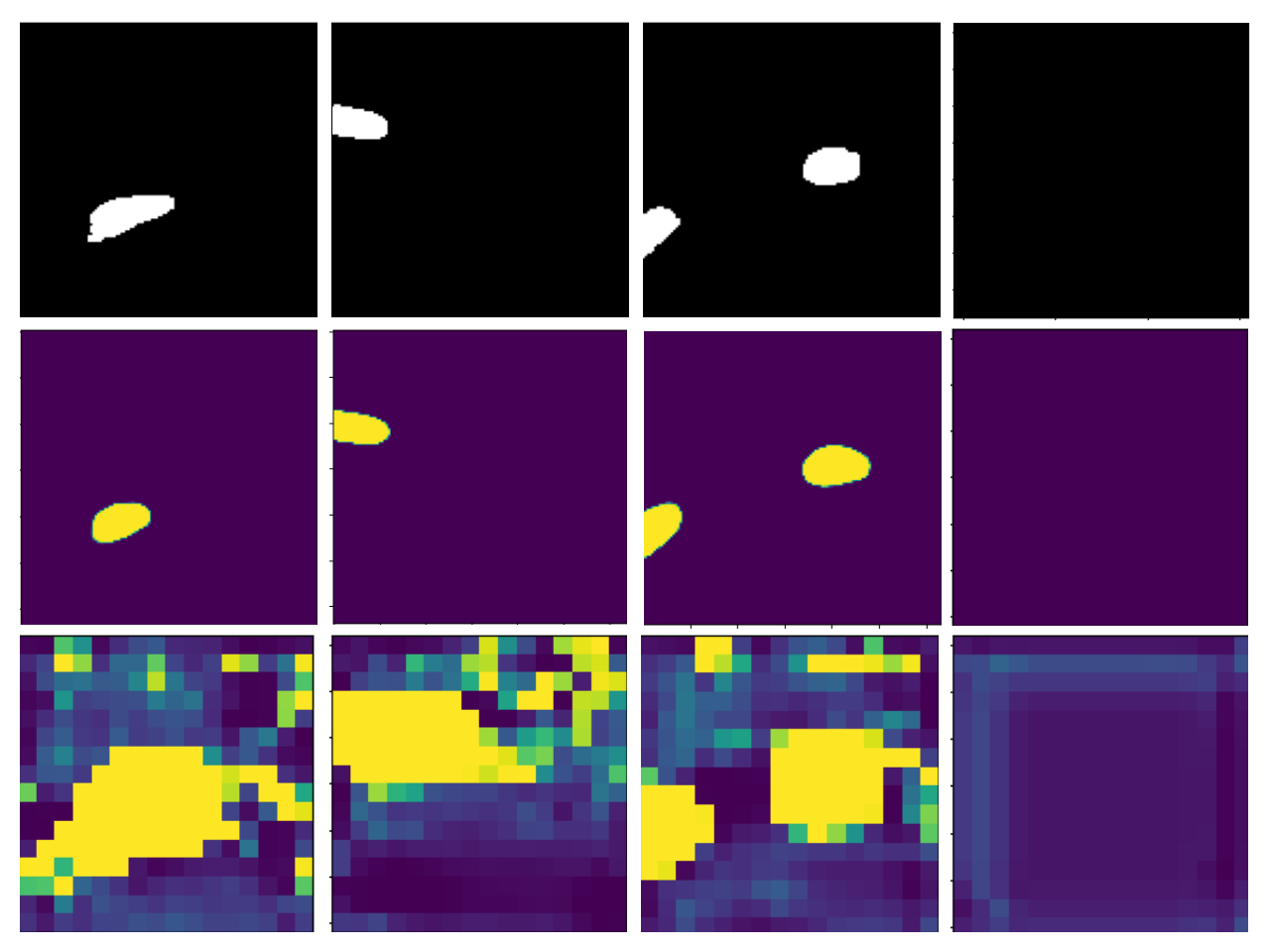}
    \caption{Visualization of results: The first row displays the ground truth masks, the second row shows the class activation maps, and the third row presents the predicted heat maps.}
    \label{fig:cammaps}
\end{figure}

The Class Activation Mapping (CAM) module's activation map guides the segmentation process, especially with only image-level annotations available. These maps highlight crucial regions for patch classification by taking a weighted mean of feature maps from the last convolutional layer. The weights are derived from the parameters of the final linear layer. Applying an activation function produces the final map, revealing active regions—such as fracture areas—in the input image. Fig. \ref{fig:cammaps} shows these activation maps, aiding in understanding how CAM highlights fracture-related regions. This helps the segmentation model identify these areas using bottleneck features.

We carried out an in-depth comparative analysis of our network against several well-established segmentation architectures in the field: UNet \cite{ronneberger2015u}, UNETR \cite{hatamizadeh2022unetr}, TransUNet \cite{chen2021transunet}, Attention U-NET \cite{oktay2018attention}, and FracNet \cite{jin2020deep}. This comprehensive evaluation was performed on publicly available datasets, ensuring consistency in pre-processing and post-processing methodologies. Considering the elongated shape of rib fractures, traditional labeling with bounding boxes may overlook crucial details, and Fracnet\cite{jin2020deep} has formulated detection as a segmentation task. The analysis results, outlined in Table \ref{tab:results1}, provide insights into the detection and segmentation performance of each model, measured using metrics like the Free-Response Receiver Operating Characteristic (FROC) \cite{bandos2009area} and Dice coefficient \cite{Yu2012Fuzzy}. The detection score is calculated as the average probability of each voxel being a fracture within a connected region. Free-response Receiver Operating Characteristic (FROC) is an evaluation metric that considers sensitivity and false positives. Sensitivity is the ratio of true positives to the total number of proposals. False positives are an average number of false positives per image and can be defined as the ratio of false positives to total CT volumes. FROC performance regarding sensitivities at specific false positive (FP) levels is reported. Specifically, the average sensitivity is calculated at FP rates of  0.5, 1, 2, 4, and 8. In image segmentation, the Dice Similarity Coefficient (Dice coefficient)\cite{Yu2012Fuzzy} is a metric used to quantify the similarity between the predicted segmentation and the ground truth segmentation of an image. The DSC for  $\text{i}^\text{th}$ sample can be calculated using the equation below.
\begin{equation}
{\text{DSC}} = \frac{2 \sum_{\text{v=1}}^{\text{N}} \text{g}_\text{iv}\text{s}_\text{iv} + \epsilon}{ \sum_{\text{v=1}}^{\text{N}} \text{\text{g}}_{\text{iv}} + \sum_{\text{v=1}}^{\text{N}} \text{s}_\text{iv} + \epsilon}
\end{equation}
where $\text{g}_\text{iv}$, $\text{s}_\text{iv}$ represents ground truth and prediction of $\text{v}^\text{{th}}$  voxel of $\text{i}^\text{{th}}$ CT volume, after post-processing and, where $\text{N} = \text{W} \times \text{H} \times \text{D}, $ where W, H and D represent width, height and depth of CT volume and $\epsilon$ represents a constant. 

The results of the Fracnet \cite{jin2020deep} are reported with only publicly available data and provided post-processing steps, \cite{jin2020deep} use additional in-house data and intense post-processing steps which are not made publicly available. Our network demonstrated significant improvement, achieving an overall dice score enhancement of 1.846\% compared to the baseline models. For a visual representation of these improvements, the segmentation performance of different baseline models is visually depicted in Fig. \ref{fig:results}. The visual results show that our model outperforms the other state-of-the-art models for fracture segmentation, especially for sample 464 and corresponding slice number 278, where the fracture region is minuscule, the model is able to identify and segment the fracture.

\subsection{Ablation Study}

\begin{figure}[]
    \centering
    \includegraphics [width=\linewidth, height=10cm]{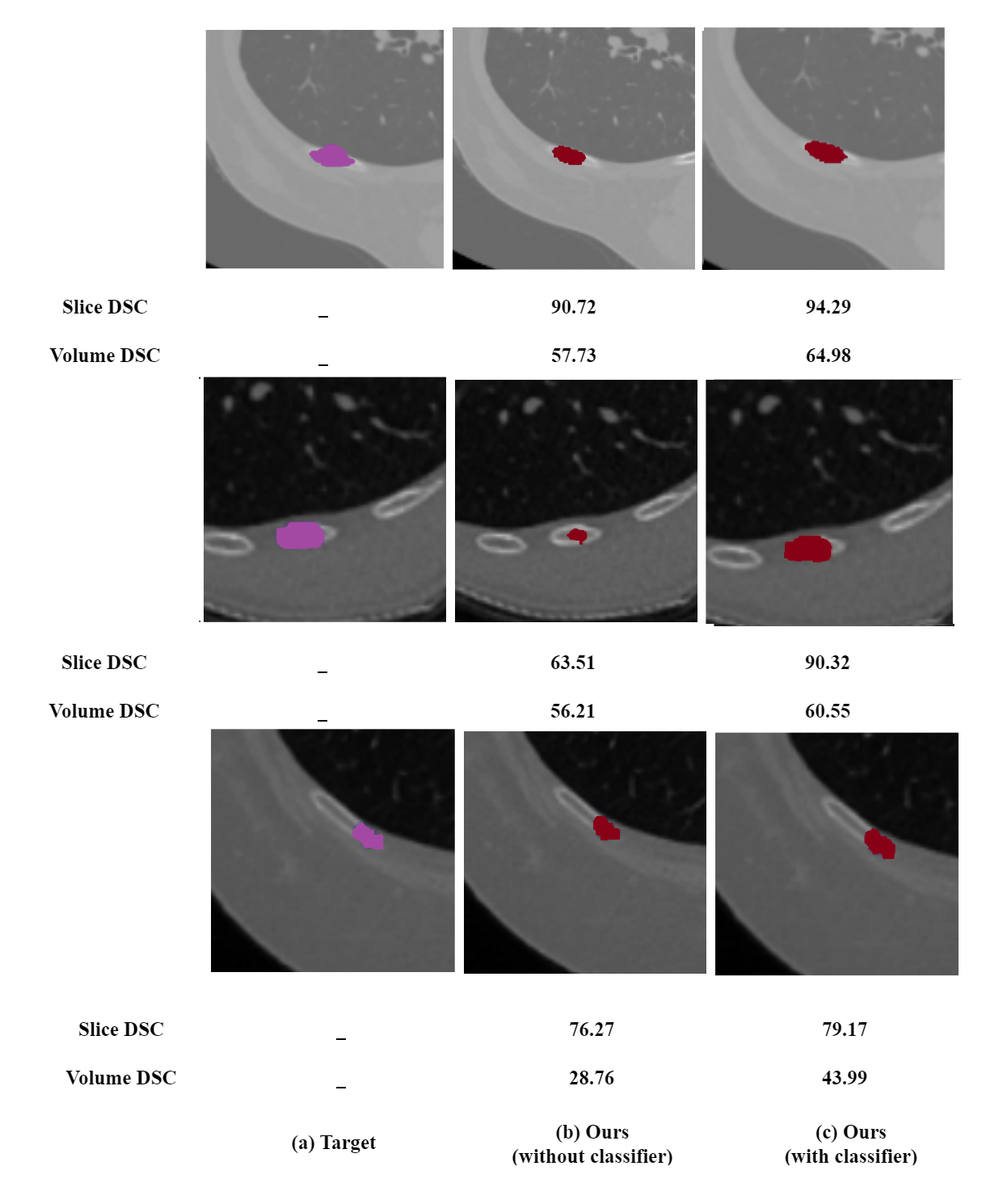}
    \caption{ Ablation results comparison with volume and slice DSC: First row: Segmentation of Sample 431, Slice Number 268, Second row: Segmentation of Sample 440, Slice Number 159, Third row: Segmentation of Sample 464, Slice Number 278.}
    \label{fig:results2}
\end{figure}

To evaluate the impact of incorporating the auxiliary classifier, we conducted training sessions with and without its inclusion to analyze its effect on segmentation results. We aimed to investigate the classifier's impact on the models' overall performance. Upon integrating the Class Activation Mapping (CAM) module into the training process, we observed overall improvement in segmentation outcomes. The network without classifier is simple UNET with Dice and focal loss, and the number of parameters used is 1401377, whereas the proposed model uses 1401506 parameters. Table \ref{tab:results1} provides a comparison between models with and without a classifier based on their performance measured by Dice Similarity Coefficient (DSC) and Free-response Receiver Operating Characteristic (FROC) scores. The architecture without the classifier uses the standard UNET but with Dice and focal loss. Specifically, there was a noteworthy increase in the Free-response Receiver Operating Characteristic (FROC) score at levels 2, 4, and 8, with improvements of 2.5\%, 2.9\%, and 2.8\%, respectively.
Additionally, the Dice Similarity Coefficient (DSC) saw a significant enhancement of 2.3\%. An additional linear layer with 129 parameters has given improved performance than the models like UNETR \cite{hatamizadeh2022unetr}, TransUNet \cite{chen2021transunet}, Attention U-NET\cite{oktay2018attention}, whose total parameters are much higher. However, it's worth noting that the improvement at FROC level 1 was marginal. Interestingly, at the 0.5 FROC level, the model without the auxiliary classifier outperformed the proposed model by 1.5\%, suggesting a higher sensitivity in the former. This indicates that while the classifier integration generally boosted performance, there are specific sensitivity thresholds where its absence might yield better results. Fig. \ref{fig:results2} presents the predicted segmentation masks of both models to illustrate an auxiliary classifier's impact visually. 

Table \ref{tab:results2} analyzes the proposed model at different probability thresholds and size thresholds, described in the post-processing step, to reduce the false positives, and the bone threshold is kept at 300 HU. The improved performance is observed at size threshold 150 and probability threshold 0.6.

\


\section{Conclusion}
Our study introduces a novel network architecture incorporating a CAM module to tackle the automatic rib fracture segmentation task. Including the CAM module facilitates improved feature discrimination at the bottleneck layer, enhancing feature representation within the model. Our experiments demonstrate that the overall performance of the proposed method surpasses the performance of other comparative methods, effectively addressing the challenge of fracture segmentation within volumes. This indicates the efficacy of our approach in segmenting rib fractures from CT volumes.
\begin{acks}
This research is funded through the grant from iHub Drishti Foundation, TIH on CV \& ARVR  under NM-ICPS, DST, Government of India.
  
\end{acks}
\bibliographystyle{ACM-Reference-Format}
\bibliography{output}

\end{document}